# Room-temperature synthesis of graphene-like carbon sheets from $C_2H_2$, $CO_2$ and CO on copper foil

M. Hajian,[a] M. Zareie,[b] D. Hashemian[a] and M. Bahrami*[ac]

[a]Faculty of Science, Isfahan University, PO Box 81746-73441, Isfahan, Islamic Republic of Iran. E-mail: moh.bahrami@ut.ac.ir

[b]Faculty of Science, University of Sistan and Baluchestan, PO Box 98135-674, Zahedan, Islamic Republic of Iran

[c]Faculty of Chemistry, College of Science, University of Tehran, PO Box 14178-63177, Tehran, Islamic Republic Iran

## Abstract

A high temperature for some catalytic reactions, like synthesis of large area and high quality graphene, is required. The mentioned graphene can be obtained by a chemical vapour deposition (CVD) process on copper foil at 800–1000 °C. Here, we describe a room-temperature synthesis method from different carbon sources for the first time, including acetylene, carbon dioxide and carbon monoxide, on copper foil by using a heuristic method, which was inspired from the role of some electronic promoters in catalyst science. Promoters are substances that increase the catalytic activity, but they are not catalysts by themselves. In this study, we used charges to modify the electronic effects of the catalysts, which were produced by piezoelectric materials.

## Introduction

There are many methods for synthesizing grapheme,[1–5] Among these, the chemical vapor deposition (CVD) method from gaseous carbon sources on catalytic metal substrates such as Ni and Cu foil has already shown great potential for large-scale graphene growth.[1,2] However, the CVD process requires a high growth temperature, typically 800–1000 °C. A low-temperature growth method is desirable because it is more convenient, economical, and feasible for industrial applications. Here, we describe the room temperature synthesis of graphene-like carbon sheets on copper foil from three stable gaseous molecules including $C_2H_2$, $CO_2$ and CO by using a heuristic method. The main part of our heuristic method has been inspired from the role of electronic promoters in catalyst science.

Promoters are the subject of great interest in catalyst research due to their remarkable influence on the activity and selectivity of industrial catalysts. Many promoters are discovered serendipitously; but few are the result of systematic research. This sector of catalyst research is often the scene of surprising discoveries.[6] There are many examples that promoters are used for increasing of catalytic activity. The ammonia synthesis is a well-known example. The promoter potassium facilitates the dissociation of chemisorbed $N_2$ on the iron and thus increases the rate of formation of $NH_3$. The strongly electropositive potassium provides electrons that flow to the metal and then in to the chemisorbed nitrogen molecule.[6–8] As Fig. 1a shows these electrons through iron catalyst are transferred into the antibonding $\pi^*$ orbitals and in this way $\pi$ backbonding into the $\pi^*$ orbitals of the adsorbate is considerably strengthened. These actions reduce the bond order of $N_2$ molecules and facilitate the dissociation of the chemisorbed molecules ($N_2$) on the catalyst.

Inspired by the role of potassium, we decide to use external static electricity as a supplying source of electrons for cleavage or weakness of chemical bonds of adsorbed molecules on catalyst surface. Static electricity is a suitable form of electric charges at rest that located over the outer surface of catalyst after charging, where the adsorption and catalytic reactions can be taken place. Also it is inexpensive, available anywhere and

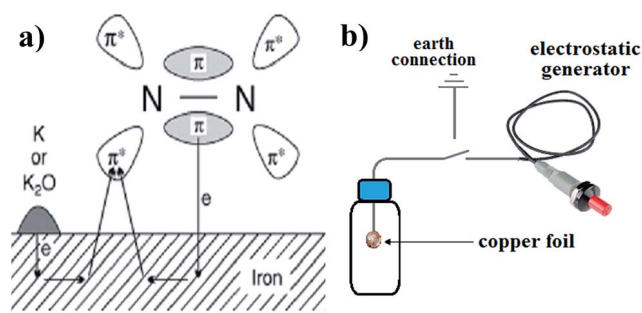

Fig. 1 (a) The action of potassium promoters in the dissociative chemisorption of $N_2$ on iron catalysts.[6] (b) A simplified schematic scheme of the experiment.

Corresponding Email: moh.bahrami@ut.ac.ir

can be discharge more easily by earth connection. When static electricity is transferred to the catalyst, the static charges remain for some time before they gradually leak away. However, the surface becomes negatively charged so that the adsorption of further molecules requires more and more energy. Therefore, the discharging process for continuing of adsorption is necessary after every charging. When the charge density of surface increases the physisorbed molecules repelled out from surface due to weak interactions. Since covalent interactions are strong enough to make the molecules stick to the surface, the

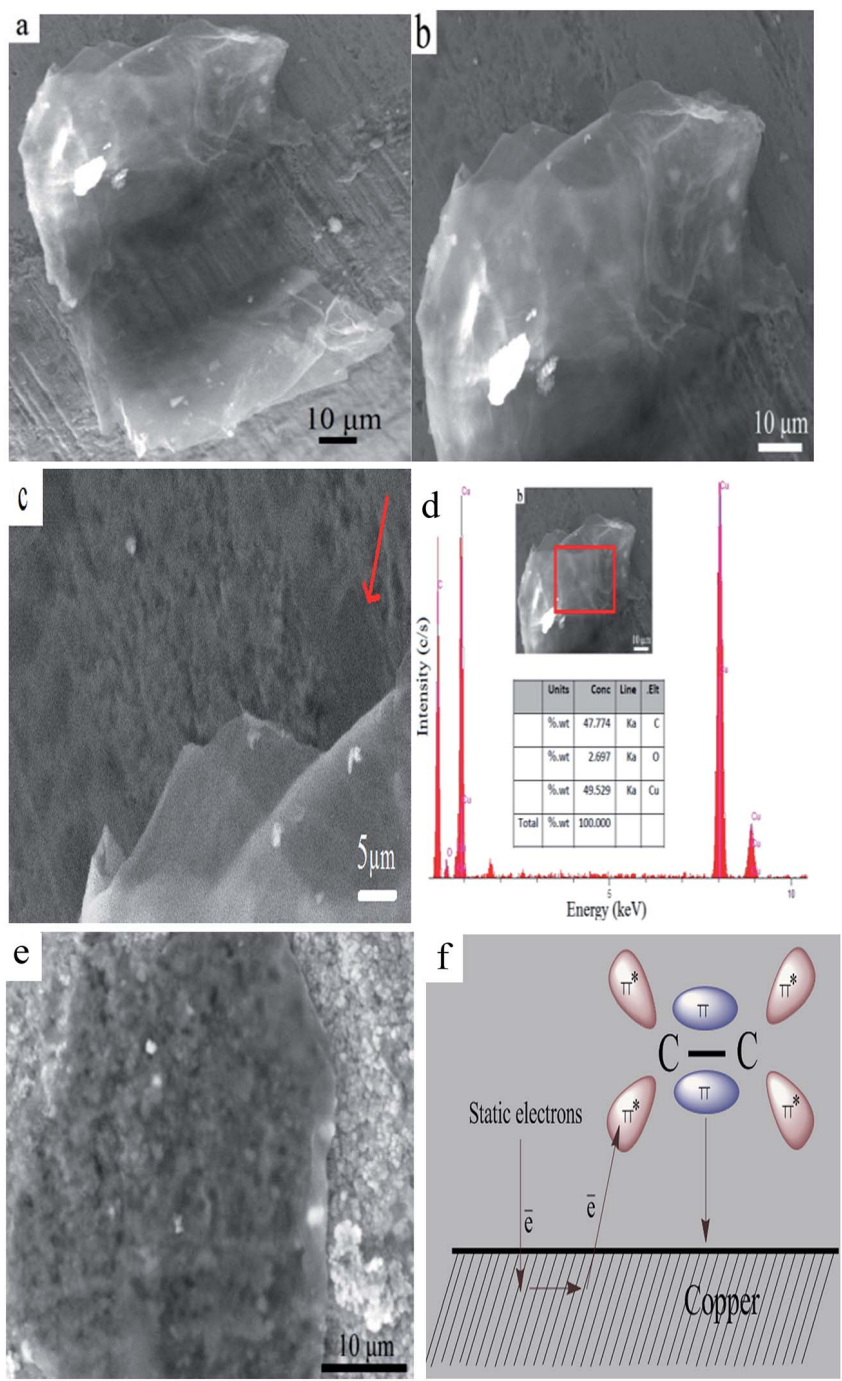

Fig. 2 (a and b) Low magnification SEM images of as-grown acetylene-derived graphene synthesized at room temperature on copper foil. (c) High magnification SEM image showing finite size graphene sheets in the form of dark irregularly shapes (one is indicated by the red arrow). (d) EDS elemental microanalysis of micron size sheets shows that they are comprised of carbon atoms. The oxygen probably comes from the copper oxide species on surface or absorbed oxygen and water molecules on graphene.[11,12] (e) SEM image of acetylene-derived graphene synthesized on copper–copper oxide foil at room temperature. (f) The action of static electrons in the decomposition of chemisorbed acetylene on copper foil. Note that all SEM images were captured without any Au coating and therefore graphene sheets represent dark contrast.

chemisorbed molecules remain on the surface. As a result, the molecular status of adsorbates on surface must be chemisorbed in this method. We tested our heuristic procedure using acetylene gas and copper foil as the first adsorbate and the metal catalyst, respectively. Acetylene is chosen for two reasons: first, it has a molecular structure similar to nitrogen molecule and second, acetylene molecules are adsorbed molecularly with C–C axis parallel to the copper surface at room temperature. They give up two π electrons of its triple bond and form two σ bonds with Cu atoms.[9,10]

## Method

The experiments were done at ambient conditions in PET bottle with volume 250 $cm^3$, containing starting material gas (including acetylene $C_2H_2$, carbon dioxide $CO_2$ and carbon monoxide CO), plastic coated copper wires and copper foils. Copper wires with length 10 cm and 1 mm in diameter were selected for electron transmitter. One end of wires was connected to a round copper foil with 0.5 cm diameter as catalyst and placed in the bottle. Other end was connected to the static electricity generator. In this work we used piezoelectric material (Piezoelectric igniter) to generate static charges. Then, carbonaceous gas was introduced into the bottle until its pressure reach to 0.5 atm. The experiments were done for 1 h in which the period times of charging and discharging process were 0.5–1 second. Note that the discharging process is occurred and strengthened due to formation of positive and negative poles. We designed this experiment so that there is no the pair of positive-negative poles around the copper foil and discharging process takes places in the region away from the copper foil by earth connection immediately after every charging. Fig. 1b shows a simplified schematic scheme of our experiment.

## Results and discussion

The SEM images of copper surface after experiment show the presence of micron-size graphene-like sheets, surprisingly (Fig. 2a–c). As can be seen in Fig. 2a–b, the edges of the large these sheets separated from the surface. Graphene is an electrical conductor; and it can be assumed that the external charges are transferred from copper foil into graphene. This situation makes reciprocal repulsive interactions between catalyst and the graphene sheet, which allow the as-grown graphene sheet to separate from the surface while small graphene remain on copper surface (Fig. 1c). EDS elemental microanalysis (Fig. 1d) also reveals that apart from the Cu signal, which comes from the copper surface, as-grown materials contain 94.7% C and only 5.3% O (the oxygen content probably arises from the copper oxide species or absorbed oxygen and water on graphene surface).[11,12] Synthesis of graphene-like carbon sheets at room temperature is surprising because its growth involves the decomposition of carbon source over a substrate typically held at 800–1000 °C.

A reasonable explanation for this astounding process is that the static electrons, generated from piezoelectric material, flow to the copper foil and suddenly increase the charge density of the catalyst surface. Subsequently this density flows from the catalyst into the antibonding orbitals of acetylene molecules and lowers the bond order (Fig. 2f), the same as dissociative chemisorption of $N_2$ on iron catalysts. According to chemistry, reduction of bond order means that dissociation can more readily occur. Thus, the influence of static charges makes that acetylene molecules completely or partially dissociate to their constituent atoms or molecular fragments (radicals or ions) on copper surface. Finally, these high reactive intermediates react rapidly to produce compounds that are thermodynamically stable at room temperature e.g. graphene-like sheets (flakes).

The growth of multilayer and suspend graphene (Fig. 2a–b) indicates that the graphene is also self-growth in this process and capable to act as a catalyst. It was understood that the edges of graphene are not well controlled in experiments, and it is hard to obtain saturated edges without any dangling bond (DB).[13] Acetylene molecules are strongly chemisorbed around the DB sites because these sites are very chemically reactive.[14] Then, acetylene decomposition can occur through the mentioned mechanism and graphene grow horizontally by the addition of carbon atoms (or species) to its edges. Thus, the graphene could grow with a similar behavior to carbon nanotube that implies the possibility for the continuous growth of large-area suspended graphene by this method. This assumption is also confirmed by the growth of graphene-like sheets on copper–copper oxide foil under same conditions, where there are the gaps of hundreds of nanometers between nearby copper oxide grains (Fig. 2e). Fig. 2e indicates that the synthesized graphene-like sheet is semitransparent with several microns in size so that the copper oxide grains are visible behind it. Some scientists believe that there is a weak interaction between graphene and the Cu substrate allows the flakes to expand over the grain boundaries.[15] It seems that the graphene nucleate on the copper surface and then dangling bonds also contributed to its growth. It is also possible that electric fields produced from external charges on the surface to be effective in dissociations and reactions mechanism. However, the precise mechanism will require further investigation using advanced methods and we take them to be beyond the scope of this paper and a subject of future researches.

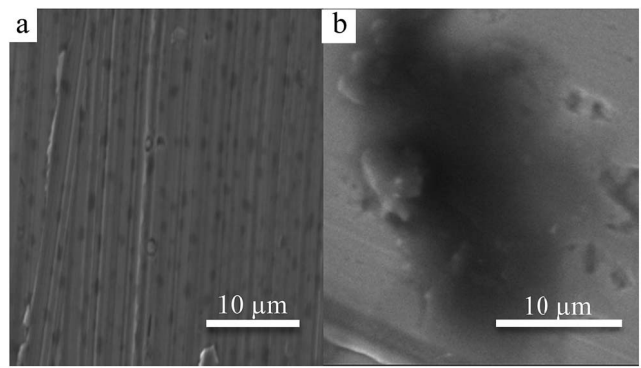

Fig. 3 SEM images of graphene-like on Cu with different growth times of (a) 30 min and (b) 45 min.

The large number of small carbon sheet domains synthesized at room temperature on copper foils which shown in Fig. 2c. We grew films on Cu foil as a function of time under same conditions. SEM images of graphene-like sheets on Cu with different growth time are shown in Fig. 3. Surprisingly, the obtained product is similar to the evolution of the graphene film deposited on copper at 800–1000 °C described by SEM images. Furthermore, it is expected that the synthesized graphene is free of wrinkles or distortions. The mentioned undesirable properties are produced due to thermal expansion coefficient difference between copper foil and graphene during cooling down in typical CVD process. Therefore, our suspended-like and smooth graphene sheets are ideal samples for the investigation of the intrinsic properties of graphene and can be contributed to the future progress of graphene technology.

However, a remaining issue is structural analysis and quantifying possible defects in as-grown graphene-like materials on substrate by this method. Although these evidences strongly indicate that this carbonaceous materials probably are graphene, but it cannot be with certainty said unless this materials characterized by atomic resolution TEM and STM. Unfortunately, such instruments do not exist available due to sanctions in our country, but this analysis can be performed and reported by other researchers.

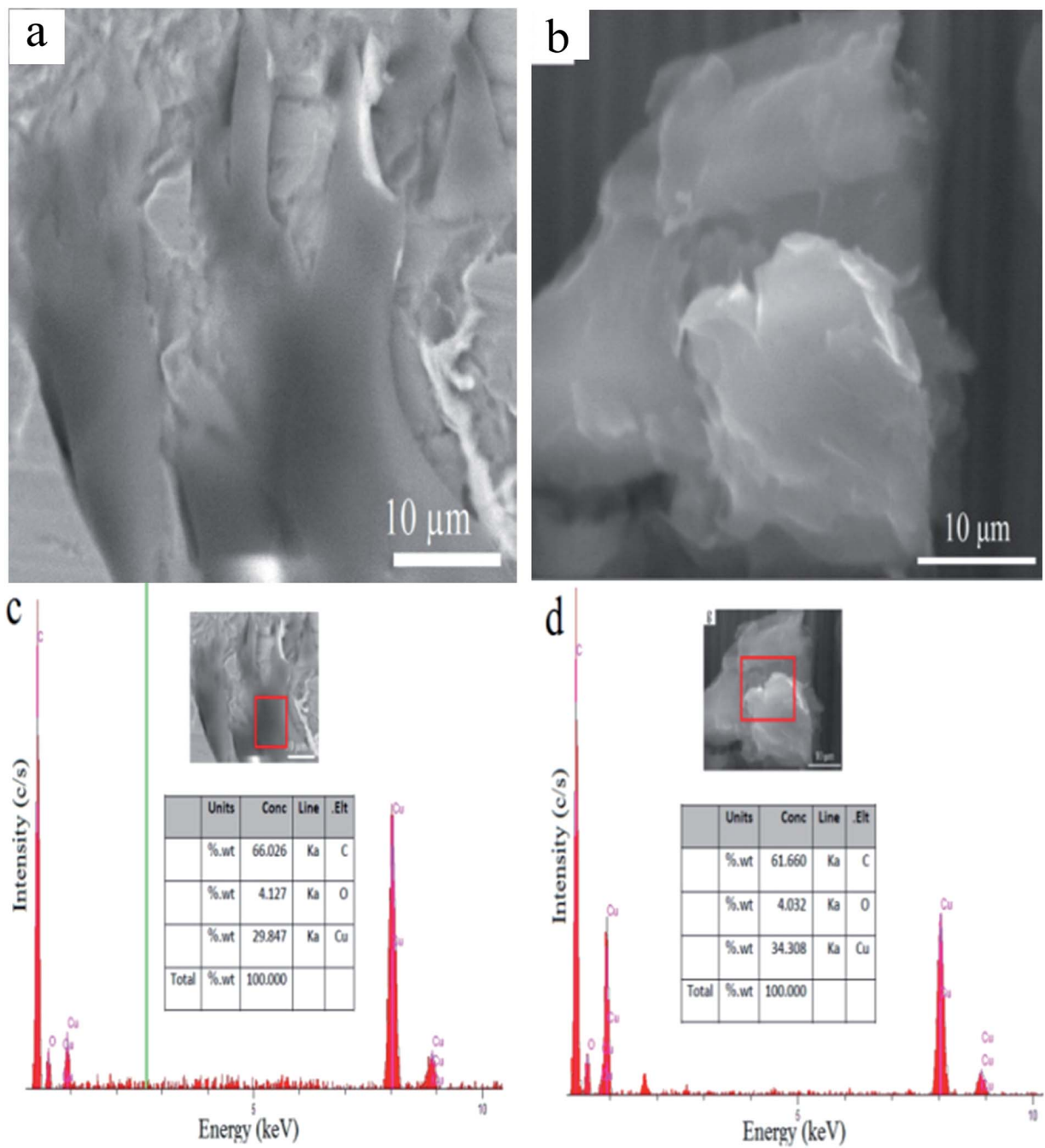

Fig. 4 SEM images of as-grown graphene flakes synthesized from (a) carbon dioxide $CO_2$ and (b) carbon monoxide CO gases on copper foil at room temperature. EDS elemental microanalysis of micron size graphene flakes derived from (c) carbon dioxide $CO_2$ and (d) carbon monoxide CO.

We were also curious to know whether this procedure can be used to decompose other stable molecules including carbon dioxide $CO_2$ and carbon monoxide CO because these molecules can be chemisorbed on copper.[16,17] We explored the possible growth of graphene-like carbon sheets on copper foil under same conditions using $CO_2$ and CO gases as starting material. The SEM images of the copper surfaces after experiments indicated that these two carbon sources were unbelievably transformed into micron-sized graphene-like flakes (Fig. 4a–b). EDS elemental microanalysis shows that as-grown materials contain 94.1% C and 5.9% O for $CO_2$ (Fig. 4c) and 93.9% C and 6.1% O for CO (Fig. 4d). The oxygen content in all flakes is the same approximately and comes from copper oxides, absorbed oxygen and water or trapped molecules on graphene surface.[11,12] The results clearly indicate high concentration of carbon and confirm that all obtained sheets are graphene-like carbon flakes. This is really exciting because carbon dioxide is chemically most stable molecule and its decomposition to carbon is thermodynamically unfavorable. Also the carbon monoxide contains strong triple bond that is very difficult to break.

## Conclusions

The main conclusion of this work concerns the room-temperature synthesis of graphene on copper foil from different carbon sources using external charges. These charges come from piezoelectric material and modify the electronic effects of copper foils, and thereby, the high stable molecules such as carbon dioxide are transformed into graphene flakes at room temperature. This work may increase the knowledge of chemical bonds and improve catalytic operation.

## Author contributions

M. Zareie, D. Hashemian and M. Bahrami discovered the method, designed and carried out experiments. M. Hajian was supervisor and financial support.

## Acknowledgements

The authors thank Mohammad H. Keshavarz; Nafiseh Alfi and J. poursafar for technical assistance. They also contributed to writing the paper.